\documentclass[12pt]{article}
\usepackage{amsmath}
\usepackage{amsfonts}
\usepackage{amssymb}
\usepackage{hyperref}
\usepackage{authblk}
\usepackage{graphicx}
\usepackage{subcaption}
\usepackage[english]{babel}
\usepackage[usenames]{color}
\usepackage{booktabs}

\title{Estimates of the Dynamic Characteristics of Binary Systems for Traversable Wormholes Search}
\author[]{}
\author{I.A.~Moiseev$^{1a}$, O.S.~Sazhina$^{2b}$}
\date{\vspace{-5ex}}

\author{
     I.A.~Moiseev$^{1, 2}$\thanks{E-mail: lxyniti@gmail.com}, O.S.~Sazhina$^{2}$\thanks{E-mail: cosmologia@yandex.ru}
 }

 \date{
     {\small
     $^{1}$Lomonosov Moscow State University, Moscow, Russia\\%
     $^{2}$Sternberg Astronomical Institute, Moscow, Russia\\[1ex]}
    \vspace{0.5em} 
     \today
 }

\begin{document}

\maketitle

\begin{abstract}

The paper presents simulated detectability of traversable wormholes (WHs) in wide binary astrophysical systems without accretion (dormant binary systems) and in the Galactic center. The wormhole spacetime is represented as a gluing of two asymptotically flat Schwarzschild spacetimes, without specifying the detailed geometry of the throat. The gravitational perturbation across the throat is calculated by matching the monopole perturbations on the two sides in the short-throat approximation. We consider a traversable WH candidate as a compact object in a binary system, whose companion star is perturbed by a massive object on an elliptical orbit on the other side of the WH throat.
In the case of a supermassive BH as WH candidate the perturbing acceleration is analyzed and compared with the stochastic gravitational influence of surrounding stars. For dormant wide binary systems, we analyze the radial velocity perturbations and discuss how their signatures may differ from those of a triple system.  We discuss the observational limits of our model arising due to the Keplerian orbit subtraction. Under the model assumptions and the considered observational sampling, radial velocity perturbations are detectable in simulations with an observational accuracy of 1.5 km/s for all considered observation time spans. For a more conservative accuracy of 5 km/s, the non-parametric analysis method becomes statistically significant after $\approx17$ years of simulated data accumulation. The spectral and non-parametric methods significantly reduce the required accumulation time compared to matched-filtering applied in isolation.

\end{abstract}

\section*{Introduction}
\noindent

Due to the increase in observational accuracy and the development of the gravitational-wave observation channel, such hypothetical objects as wormholes (WH) are becoming not only objects of active theoretical studies, but also objects of observational search. 

WHs can arise as solutions of classical GR theory and a number of modified gravity theories and represent a hypothetical space-time tunnel that is fundamentally capable of connecting two remote regions of the Universe, or, in the case of Multiverse models, connecting different universes. 

Assuming that WHs can exist in the Universe along with reliably and massively detected population of black holes (BH) (in our Galaxy, the estimated number of BHs is on the order of $10^8$, \cite{Sahu}), designing concrete observational programs becomes particularly important. To do this, it is necessary to identify and quantify potentially observable characteristics and properties of WHs that may differ significantly from those of BHs.

Let us note a number of papers devoted to the search for possible astrophysical WH manifestations and their differences from BH: \cite{Zaslavskii}--\cite{Karimov}. The approaches include, among other things, a comparative analysis of the effects of gravitational lensing and the formation of WH and BH shadows, the study of anomalies in the star motion in the immediate vicinity of the proposed WH candidates, the analysis of the characteristics of the spectra of accretion disks and gravitational wave radiation in binary systems with WH candidate and a companion star.

\begin{figure}
    \centering
    \includegraphics[width=0.6\linewidth]{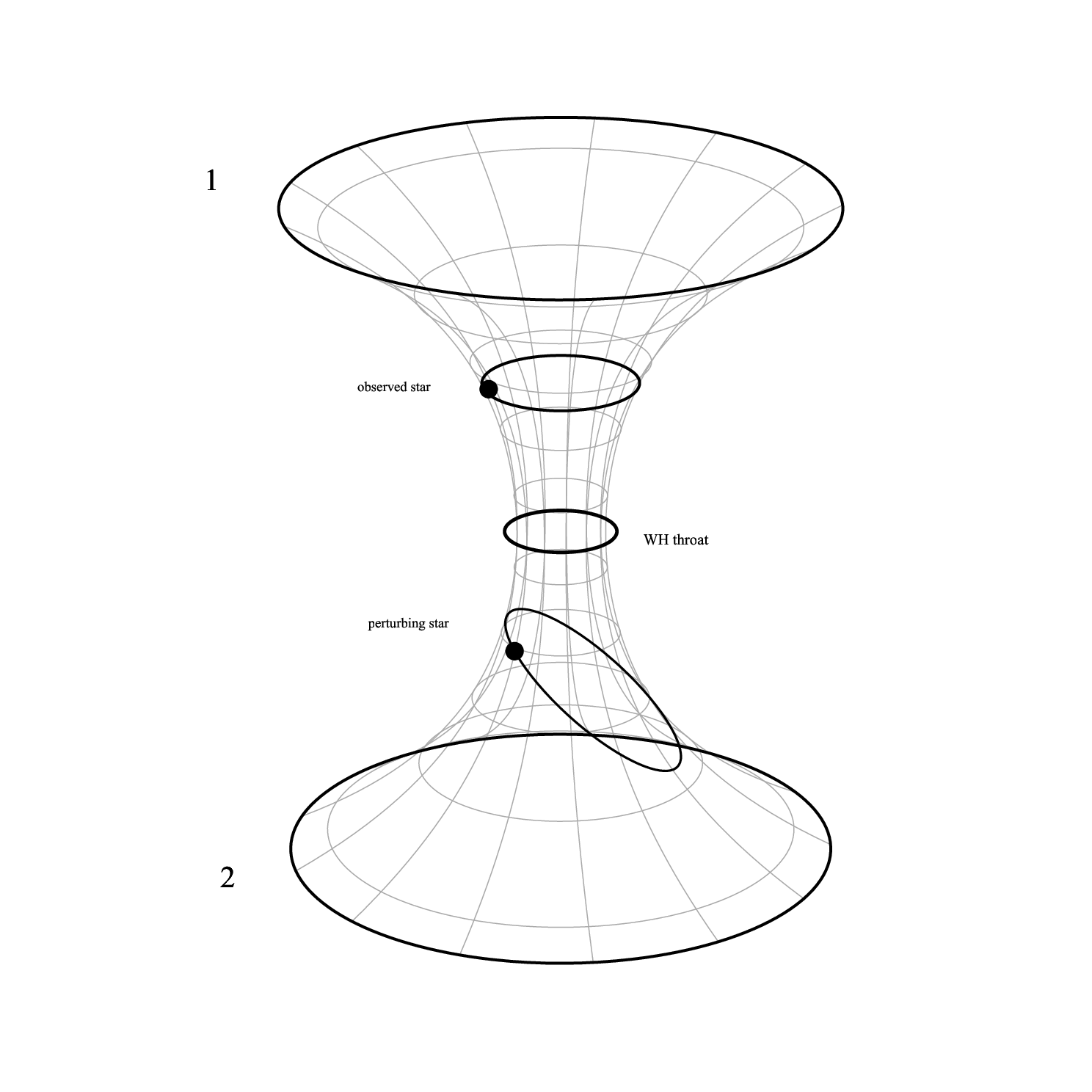}
    \caption{Schematic of the system model. The observed star orbits on side 1; the perturbing star orbits on side 2, connected to side 1 via the WH throat.}
    \label{fig:model}
\end{figure}

Among these studies, the theoretical development and the search for observational manifestations of traversable WH are of particular interest. The importance of this approach is due to the following factors. 
\begin{itemize}
\item Recent DESI results provide interesting cosmological hints towards evolving or phantom-like dark energy \cite{Berti}--\cite{Teixeira}, which provides a motivation for considering phantom energy as a possible source of the exotic matter required to stabilize the throat of a traversable WH.
\item In contrast to studying the effects caused by the difference between the WH and BH metrics that occur in the immediate vicinity of WH candidates, in the case of traversable WHs, it becomes possible to consider wide systems that do not depend on a specific type of metric. In this approach, the gravitational influence of massive objects located on one side of the WH throat (hereinafter referred to as ``side 2'') on objects located on the other side of the WH throat (i.e. from the observer's side, hereinafter referred to as ``side 1'') is studied, see Fig. (\ref{fig:model}). For the wide astrophysical systems, corrections related to the detailed metric structure are expected to be subdominant.
\end{itemize}

Traversable WHs, as the solutions of Einstein equations, were obtained in 1973 by Bronnikov \cite{Bronnikov} and, independently, Ellis \cite{Ellis}, but they became widely known after the two famous papers by Morris and Thorne \cite{Morris 1}--\cite{Morris 2} and Visser \cite{Visser}. The theory of traversable WH  was further studied in a number of works by \cite{Bronnikov 2} --\cite{Moiseev 3}, and see references therein.

Among the most significant effects of objects on side 2 on objects on side 1 are the following.
\begin{itemize}
\item In the presence of a time-varying gravitational source on side 2 (a star with an orbit with nonzero eccentricity), the object on side 1 will be subject to gravitational perturbation, which will cause an additional acceleration determined by both the parameters of the perturbing object and the WH effective mass.  
\item The gravitational influence of the object on side 2 should produce a systematic perturbation of the radial velocity of the object on side 1, which is cumulative and can be detected by statistical analysis methods.
\end{itemize}

Earlier in the work \cite{Moiseev 1} it was shown that when considering a supermassive BH in the center of our Galaxy SgrA$^*$ as a WH candidate, and stars in its vicinity as observed objects, the magnitude of the perturbing acceleration was $a\sim 10^{-4}$ cm/s$^2$  for the S2 star, which has the greatest observational prospects. This paper also considered the main competing effects: the influence of the dark matter halo and the influence of the background of surrounding stars, not accounting for the stochastic influence of stars, which will be considered in the present paper. It was also shown in \cite{Moiseev 1} that for wide binary star systems, assuming the presence of a traversable WH and a companion star, the magnitude of the perturbing acceleration varies from $a\sim 10^{-4}$ to $a\sim 10^{-2}$ cm/s$^2$, which is already comparable to the current accuracy of acceleration measurement. The following objects were considered as wide binary systems: BH1\cite{11}--\cite{12}, BH2\cite{2}, BH3\cite{3}. Additional studies are needed to assess the effects arising from the presence of a traversable WH, namely, the analysis of changes in radial velocities and estimates of the corresponding time periods when the effect may be noticeable, which is necessary for planning future observations.

It is important to note that the effect we consider is based on the assumption, adopted in \cite{Moiseev 1}, that objects (stars) located on opposite sides of the WH throat can interact gravitationally with each other. The transmission of gravitational perturbations, as well as electromagnetic signals and other possible signals (including scalar fields that may arise in extended theories of gravity) through the WH requires the throat to be traversable and stabilized. Therefore, we consider only traversable WHs. For non-traversable WHs, the cross-throat interaction mechanism considered in the paper would not operate, and the model developed here would therefore not be applicable. Distinguishing such objects from BHs would therefore require different observational tests, for example, analyses of their shadows or other signatures associated with their accretion disks, which is not included in the scope of the present work.

The article is organized as follows.

In Section 1, the stochastic effect of stars is considered as a competing effect that can contribute to the acceleration of a star. Following the paper \cite{Moiseev 1}, we consider the star S2 and assume that the object identified with the supermassive BH in the center of our Galaxy is a traversable WH. On the other side of the throat (on side 2), we assume a massive object in orbit with a non-zero eccentricity. This completes the consideration of the main effects that can produce the perturbing acceleration as alternatives to the traversable WH hypothesis.
We also note here that the estimated stochastic influence of stars is significant only for systems containing a supermassive BH in the centers of galaxies as a candidate for a traversable WH. For binary star systems, this effect is negligible due to the lower density of stars in the vicinity of such systems.

Furthermore, the paper considers only binary star systems that are wide in order to neglect the accretion of the companion star's matter into a compact object (WH candidate).

In Section 2, a model of a traversable WH located in a wide binary star system is considered. Radial velocity anomalies and their stochastic component are considered. We construct a template to detect or rule out the presence of the perturbation in the radial velocity data. We construct the SNR statistic, which indicates the relation between the observed integral signal and the characteristic scale of noise fluctuations.  

In Section 3, simulations of the perturbing object's effect on the radial velocity of the observed object are considered.

Section 4 discusses spectral analysis and nonparametric analysis of the model (LS periodogram, ALK statistic).

In Section 5, we examine whether the proposed WH signal can be distinguished from a Keplerian orbit when the orbital parameters are allowed to vary during the fit. Section 6 contains discussion and conclusions about the prospects of our approach to searching for traversable WH in the class of so-called dormant binary and presumably triple star systems, which is currently being actively studied.

\section{The stochastic influence of stars in the S-cluster}
\noindent


We consider the neighborhood of the Galactic center (the S-cluster) as an inhomogeneous volume with a stellar density falling off as $r^{-p}$. Following \cite{Kandrup}, the mean magnitude of the stochastic force per unit mass of a test star in such a volume can be written as:
\begin{equation}
\label{eq:F}
    \langle |\mathbf{F}| \rangle = \frac{2}{\pi} D(p) \left( \frac{B(p)}{2} \right)^{\frac{2}{3 - p}} \cdot G m \alpha^{\frac{2}{3 - p}},
\end{equation}
where the $D(p),B(p)$ and the parameter $\alpha$ are given by

\[
D(p) = \frac{5 - p}{2} \, \Gamma\left( \frac{1 - p}{2 - p} \right) - \frac{3 - p}{2} \, \Gamma\left( \frac{4 - 2p}{3 - p} \right),
\]

\[
B(p) = \int_0^{\infty} \frac{z - \sin z}{z^{\frac{7 - p}{2}}} \, dz,
\]

\[
\alpha = \frac{N_\text{field} (3 - p)}{R_{\text{max}}^{3 - p}}.
\]

 Here $p$ -- the density slope parameter, $R_{\text{max}}$ -- an arbitrarily chosen boundary of the system, $N_\text{field}$ -- the number of the stars within this boundary, $m$ -- the mass of an individual field star (all field stars are assumed to have equal masses for simplicity).

This estimate is an overestimate, since equation (\ref{eq:F})  was derived for an effectively infinite volume with $N_\text{field}\rightarrow\infty,\;R_{\text{max}}\rightarrow\infty$  is applicable when the effective size of the volume greatly exceeds the mean inter-stellar distance. 

Formulae for the distribution of the stochastic force in a finite system are obtained in \cite{Popolo},  but are given in integral form, whereas \cite{Kandrup} provides not only the distribution itself but also a closed-form expression for the mean force magnitude, reducible to standard tabulated functions and directly usable by simple substitution. It is also noted in \cite{Popolo} that for large $N_\text{field}>1000$ the distribution obtained in \cite{Kandrup}, takes the form (\ref{eq:F}). Therefore, the result of computing the stochastic force per unit mass (i.e., the acceleration) from equation (\ref{eq:F}) can be regarded as an upper bound on the stochastic perturbation.

We model the region near the Galactic center as a sphere of radius $0.03$--$0.05$ pc that contains $10^3$--$10^4$ stars with a total mass below $3000M_{\text{Sun}}$  within the orbit of the star S2\cite{S2} considered previously in \cite{Moiseev 1}.

Then the relevant parameters are as follows:
\[
R_{\text{max}} \simeq 0.05 \, \text{pc}, \quad p \simeq 0.8-0.9, \quad m \simeq 1-2 M_\odot, \quad N_\text{field} \sim 10^3 - 10^4.
\]

Substituting these values into (\ref{eq:F}), we find that the perturbing contribution is
\begin{equation}
\label{eq:F2}
\langle |\mathbf{F}| \rangle \approx (9.40\times 10^{-5} \ \text{--} \ 8.1\times 10^{-4})\ \text{cm/s}^2.
\end{equation}

As noted above, this estimate is deliberately conservative in the sense of being an upper bound. For optimal parameter choices -- low stellar masses, a flatter spatial distribution, and a smaller total number of stars in the cluster -- the estimate would fall below (\ref{eq:F2}) and would be at least two orders of magnitude below the corresponding estimates for the star S62 in the extreme case (see \cite{Moiseev 1}).

Thus, the stochastic gravitational influence of surrounding stars on the acceleration of S2 turns out to be the most significant among all competing effects considered (dark matter halo and the background stellar population). The range in Eq. (\ref{eq:F2}) extends from values comparable to the estimated WH-induced acceleration to values several times larger. The periodicity of the WH perturbation may nevertheless help distinguish it from the stochastic stellar contribution. Finally, we note that such stochastic perturbations are absent in wide binary systems of the type considered here -- BH1, BH2, and BH3.

The estimate (\ref{eq:F2})can be refined in future work by numerically evaluating the mean stochastic force:
\[
\langle F\rangle=\int F\cdot W(F) \;dF,
\]
where
\[
W(F)=4\pi F^2W(\textbf{F})
\]
with the distribution $W(\textbf{F})$ obtained in \cite{Popolo}.

The estimate obtained above completes the accounting for the major competing effects (namely dark matter halo, background stellar population, and stochastic stellar perturbations) in the vicinity of the Galactic center. For S2, the range in Eq. (\ref{eq:F2}) extends from a stochastic acceleration comparable to the estimated perturbation from a massive object on side 2 to values several times larger. Sgr A* remains a possible target for a WH search, although such a search would have to account for the stochastic stellar contribution. However, the characteristic timescale of the Galactic center dynamics in addition to the limitations of current observational accuracies for stars in the S-cluster, makes wide binary systems a considerably more promising target. Systems such as Gaia BH1, BH2, and BH3 offer better radial velocity measurement precisions and have fewer theoretically justified competing effects to the perturbation effect we consider. We therefore turn our attention to this class of systems in the following sections.


\section{Analysis of radial velocity perturbations of the observed star in a system with a WH}

\subsection{Radial velocity}
\noindent

First, we present a brief derivation of the perturbations under consideration and describe the model assumptions. The objects on sides 1 and 2 are substantially apart from the WH mouths (at distances orders of magnitude larger than the gravitational radius of a compact object with an effective mass $M$), which allows us to assume that the background is close to Schwarzschild. Therefore, the WH in this case is a tunnel that connects two asymptotically flat regions. The propagation time through the throat is assumed to be negligible compared to the orbital timescales considered here. Similarly, gravitational redshift effects associated with the throat are negligible for objects located far from the mouths, where the spacetime is approximately asymptotically flat. 

The traversability of the WH allows one to study the perturbations of the spacetime induced by the object on one side as experienced by the object on the other side. Since the background is considered close to Schwarzschild, one can study its monopole perturbations as were taken in \cite{Moiseev 1}:
\[
h_{tt} = \cfrac{2Gm}{r}H(r-R_2)+\cfrac{2Gm}{R_2}H(R_2-r);
\]
\[
h_{rr}=\cfrac{2Gm r}{(r-r_g)^2} H(r-R_2).
\]

Here $m$ is the mass of the object on side 2, $H(\cdot)$ is the Heaviside function, $h_{tt}$ and $h_{rr}$ are the spatial and temporal components of the perturbed metric when considering the standard decomposition $g_{\mu\nu}=\tilde{g}_{\mu\nu}+h_{\mu\nu}$ and $R_2$ is the position of the perturbing object. 

One can write the components of such perturbation on sides 1 and 2 as:
\[
h_{tt}^{1}(r_1) = \cfrac{2b_{tt}}{r_1}, \,\,\, h_{rr}^{1}(r_1) = \cfrac{2b_{rr}r_1}{(r_1-r_g)^2};
\]

\[
h_{tt}^{2}(r_2) = h_{tt}(r_2) + \cfrac{2a_{tt}}{r_2}, \,\,\, h_{rr}^{2}(r_2) = h_{rr}(r_2) + \cfrac{2a_{rr}r_2}{(r_2-r_g)^2},
\]

which after matching those perturbations at the the WH mouth, considering a short throat model of \cite{Moiseev 1}, yields
\[
b_{tt}=-a_{tt}=Gm \cfrac{R}{R_2}, \,\,\, b_{rr}=-a_{rr}=Gm \cfrac{R}{R_2}.
\]

In the weak field approximation one finds that $h_{tt}\sim2\Phi^\text{eff}$ and therefore the additional monopole acceleration of the object on side 1 from the presence of the object on side 2 is
\[
\vec{a} = -\vec{\nabla}\Phi^\text{eff}=Gm \cfrac{R}{R_2} \cfrac{1}{r_1^2}\cdot \vec{e}_r=-\frac{GmR}{R_2}\frac{\vec{r}_1}{r_1^3},
\]
with its absolute value
\[
a=Gm\frac{R}{R_2}\frac{1}{r_1^2}
\]

Since the elliptic orbit of the perturbing object can not be represented as a single monopole, one should consider instead the variation of the additional acceleration:
\[
\Delta a=Gm\left(\frac{R}{r_{p}}-\frac{R}{r_{a}}\right) \frac{1}{r_{1}^{2}},
\]
where $r_p$ is the pericenter radius of the perturbing star's orbit on side 2, $r_a$ is the apocenter radius of the perturbing star's orbit on side 2, $r_1$ is the radial coordinate on side 1, and $R$ is the WH throat radius.

We can express the orbital period of the perturbing object $T$ via Kepler's third law:
\[
\quad n^{2} a^{3}=G(M+m),
\]
where $M$ denotes an effective mass of the WH,
\[
\frac{4 \pi^{2} A^{3}}{G(M+m)}=T^{2}.
\]

The impulse duration is then \cite{2007.12184}: 
\[
\tau_{\text {pulse }} \sim \left(\frac{r_{p}}{r_{a}}\right)^{2} T=2 \pi\left(\frac{r_{p}}{r_{a}}\right)^{2} \sqrt{\frac{A^{3}}{G(M+m)}},
\]
where $r_p$, $r_a$, $A$, $m$ denote the pericenter radius, apocenter radius, semi-major axis, and mass of the star on side 2, respectively.

The velocity perturbation accumulated by the star on side 1 over the time $\tau_{\text {pulse }}$ is then
\[
\Delta v \sim \tau_{\text{pulse}} \Delta a = 2\pi \left(\frac{r_{p}}{r_{a}}\right)^2 \sqrt{\frac{A^3}{G(M + m)}} \frac{G m R}{r_1^2} \left(\frac{1}{r_{p}} - \frac{1}{r_{a}}\right).
\]

An additional question here arises about the WH throat radius. Since the throat radius is not fixed in our calculation, the gravitational radius (of the object with the same effective mass $M$) provides the natural characteristic length scale for the throat radius. Therefore, we substitute the following expression for the WH throat radius:
\[
R = \alpha\frac{2 G M}{c^2},
\]
where $\alpha$ is some arbitrary factor, with $\alpha>1$ corresponding to a throat radius larger than the gravitational radius, dependent on the exact metric properties, which are not in the scope of this approach. Within the adopted perturbation model, the uncertainty in the throat radius enters the expression of the perturbation only through the overall normalization of the perturbation amplitude, while its radial dependence and periodic structure remain unchanged.

We obtain the final expression for the change in the velocity of the star on side 1 due to the gravitational influence of the star on side 2:
\begin{equation}
    \Delta v = 4\pi\alpha \frac{G^2 M m}{r_1^2 c^2} \sqrt{\frac{A^3}{G(M + m)}} 
\left(\frac{r_{p}}{r_{a}}\right)^2 \left(\frac{1}{r_{p}} - \frac{1}{r_{a}}\right)\equiv\frac{const}{r_1^2}, 
\label{eq:deltav}
\end{equation}
where $r_1$ is the radial coordinate on side 1.

Expression (\ref{eq:deltav}) is obtained without explicitly accounting for the time dependence of $r_1$  which will subsequently be used in constructing a more accurate perturbation template. Thus, for $r_1=r_1^\text{pericenter}=const$ a single impulse at pericenter takes the form:
\[
\Delta v (t) = \Delta v \cdot \delta (t - t_p), \quad \Delta v = \Delta v |_{r_1=r_{1}^\text{pericenter}} = const,
\]
since for $P \gg T$ (the perturbing object has a much shorter orbital period $T$ than the observed star's orbital period $P$) the net secular velocity perturbation consists of contributions from negative and positive impulses (corresponding to the receding and approaching phases of the observed star on side 1 relative to its pericenter), which can be treated as a single impulse per period $P$. This impulse is assigned to the pericenter, since the magnitude of the velocity perturbation is inversely proportional to the square of the radial coordinate of the star on side 1. The contribution of multiple impulses at successive pericenter passages of the perturbing star on side 2 can be accounted for without modifying the structure of $\Delta v$ simply by substituting $r_1=r_1(t_m),\; t_m=t_0+mT$.

The observable quantity is the line-of-sight velocity of the star, whose vector is determined by the Keplerian velocity and the perturbation derived above (\ref{eq:deltav}):
\[
\vec{v} = \vec{v}_0 + \Delta v \cdot \vec{e_r},
\]
where $\vec{v}_0$ is the Keplerian velocity of the star and $\vec{e_r}$ is the unit vector of the perturbing force direction:
\[
\vec{e_r}(t)=-\frac{{\bar{r}}_1(t)}{\sqrt{{x}(t)^{2}+{y}(t)^{2}}}.
\]

In the last expression it is taken into account that the star's motion is integrated in a fixed plane, within which the time dependence of the Cartesian coordinates $x(t)$, $y(t)$ is defined.

To project onto the line of sight, the rotation matrix contains two orientation angles $\Omega, \,i$:
\[
R = R_z(\Omega) R_x(i),
\]
where
\[
R_z(\Omega) =
\begin{pmatrix}
\cos\Omega & -\sin\Omega & 0 \\
\sin\Omega & \cos\Omega & 0 \\
0 & 0 & 1
\end{pmatrix},
\quad
R_x(i) =
\begin{pmatrix}
1 & 0 & 0 \\
0 & \cos i & -\sin i \\
0 & \sin i & \cos i
\end{pmatrix}
\]

Thus, transforming to the observer's reference frame:
\[
\vec{v}^{observer} = R \cdot \vec{v}
\]

The line-of-sight velocity of the star in our observer-centered coordinate system is then the $z$-component of the transformed vector $\vec{v}^{observer}$:
\[
v^{o} = [\vec{v}^{observer}]_z = \left[ R_z(\Omega) R_x(i) (\vec{v}_0 + \Delta v\cdot \vec{e_r}) \right]_z.
\]

\subsection{Accounting for the stochastic component in the radial velocity}
\noindent

\subsubsection{Radial velocity noise}
\noindent

In the case of real observations, the radial velocity of the star will be subject to noise. For simplicity, we assume this noise arises solely from observational errors and can therefore be considered Gaussian.

The observed line-of-sight velocity can then be written as:
\[
v = v^{o} + n(t),
\]
where the Fourier transform of the noise $\tilde{n}(t)$ satisfies:
\[
\left\langle\tilde{n}(f) \tilde{n}^{*}\left(f^{\prime}\right)\right\rangle=\frac{1}{2} S_{n}(f) \delta\left(f-f^{\prime}\right)
\]

Therefore:
\[
v^{o}(t) = \left[ R_z(\Omega) R_x(i) (\vec{v}_0(t) + \Delta v\cdot \vec{e_r}(t)) \right]_z,
\]
\[
\vec{e_r}(t)=-\frac{{\vec{r}}_1(t)}{\sqrt{{x}(t)^{2}+{y}(t)^{2}}},
\]
\[
v(t) = v^{o}(t) + n(t),
\]

The problem is thus to find the signal template -- the line-of-sight velocity of star 1 including the perturbation from the star on side 2 -- and to construct an optimal filter for noise suppression.

\subsubsection{Template construction}
\noindent

We compute the template as follows. Consider the residual function:
\[
r(t) \equiv v^{o}(t)-v_0(t)=s(t)+n(t),
\]
where $v_0(t)$ is the Keplerian component of the line-of-sight velocity, and $s(t)$ denotes
\[
s(t) = \left[ R_z(\Omega) R_x(i) \Delta v\cdot \vec{e_r}(t) \right]_z.
\]

The expected signal in the absence of noise can be written as:
\[
s(t) =K \cdot u(t),
\]
where $u(t)$ is the normalized signal template and, $K$ is a scaling factor proportional to the perturbing acceleration $\Delta a$. This representation is particularly convenient in the realistic search scenario where the perturbation amplitude is likely to be unknown in advance, and the entire response is described by the unique impulse structure encoded in the template $u(t)$.

To determine the signal-to-noise ratio (SNR) for the observations, we apply the matched filtering procedure. Taking the Fourier transform:
\[
r(t)=s(t)+n(t) \,\, \to \,\,
\tilde{r}(f)=\tilde{s}(f)+\tilde{n}(f),
\]
and for a template as well:
\[
u(t) \,\, \to \,\, \tilde{u}(f).
\]

The correlation ($S$) of the signal with the template gives the SNR in the following form:
\[
S = \int\limits_\mathbb{R} \tilde{r}(f) \tilde{Q}^*(f) \, df,
\]
\[
\text{SNR} = \frac{\langle S \rangle}{\sigma_N} 
\]


where  $N$ is the sample size, and the specialized frequency-domain template is:
\[
\tilde{Q}(f) = K \frac{\tilde{u}(f)}{S_n(f)} e^{2\pi i f t_0}
\]

Using standard results for Gaussian noise:

\[
\sigma_N^2 = \int\limits_\mathbb{R} S_n(f) |\tilde{Q}(f)|^2 \, df,
\]
where
\[
S_n(f) = \langle |\tilde{n}(f)|^2 \rangle + \langle |\tilde{n}(-f)|^2 \rangle.
\]

For the template $\tilde{Q}(f)$ we obtain the time-shift-dependent SNR($t_0$), \cite{Allen}:
\begin{equation}
    \label{eq:SNR}
    \text{SNR}(t_0) =
\frac{ \int_0^\infty \frac{\tilde{r}(f)\tilde{u}^*(f)}{S_n(f)} e^{2\pi i f t_0} df }
{\left[ \int_0^\infty \frac{|\tilde{u}(f)|^2}{S_n(f)} df \right]^{1/2}}. 
\end{equation}

Let us now describe in detail the procedure for determining the normalized template $u(t)$. Grouping all constant quantities into a single constant and retaining the time dependence of the single-impulse amplitude:
\[
\Delta v (t) = \frac{const}{r_1^2(t)} \equiv \frac{C}{r_1^2(t)}, \quad
r_1(t) \equiv \sqrt{x^2(t) + y^2(t)}.
\]

The impulse times are defined as:
\[
t_k = t_0 + k T,\quad k \in \mathbb{Z}.
\]

The number of impulses $N_{imp}$ falling within the template interval \(P\):
\[
N_{imp} = \left\lfloor \frac{P}{T} \right\rfloor. 
\]

The raw template is:
\[
\sum_{k=0}^{N_{imp}-1} a_{k} \delta\left(t-t_{k}\right)=\sum_{k=0}^{N_{imp}-1} \frac{C \delta\left(t-t_{k}\right)}{r_{1}^{2}\left(t_{k}\right)}.
\]

After normalization and averaging, the template takes the form:
\begin{equation}
\label{eq:u}
u(t) =\frac{1}{W} \sum_{k=0}^{N_{imp}-1}\left(a_{k}-\bar{a}\right) \delta\left(t-t_{k}\right),
\end{equation}
where the auxiliary quantities  (the mean perturbation amplitude $\bar{a}$ and the normalization coefficient $W$) are:
\[
\bar{a} =\frac{1}{N_{imp}} \sum_{k=0}^{N_{imp}-1} a_{k}=\frac{C}{N_{imp}} \sum_{k=0}^{N_{imp}-1} \frac{1}{r_{1}^{2}\left(t_{k}\right)},
\]
\[
W =\sqrt{\sum_{k=0}^{N_{imp}-1}\left(a_{k}-\bar{a}\right)^{2}}.
\]

The perturbation impulse times are again defined by:
\[
t_{k} =t_{0}+k T,
\]
where the orbital period of the perturbing object follows from Kepler's third law:
\[
T=2 \pi \sqrt{\frac{A^{3}}{G\left(M+m\right)}},
\]
with $A$ and $m$ being the semi-major axis and mass of the star on side 2, respectively.

\begin{figure}
    \centering
    \includegraphics[width=0.9\linewidth]{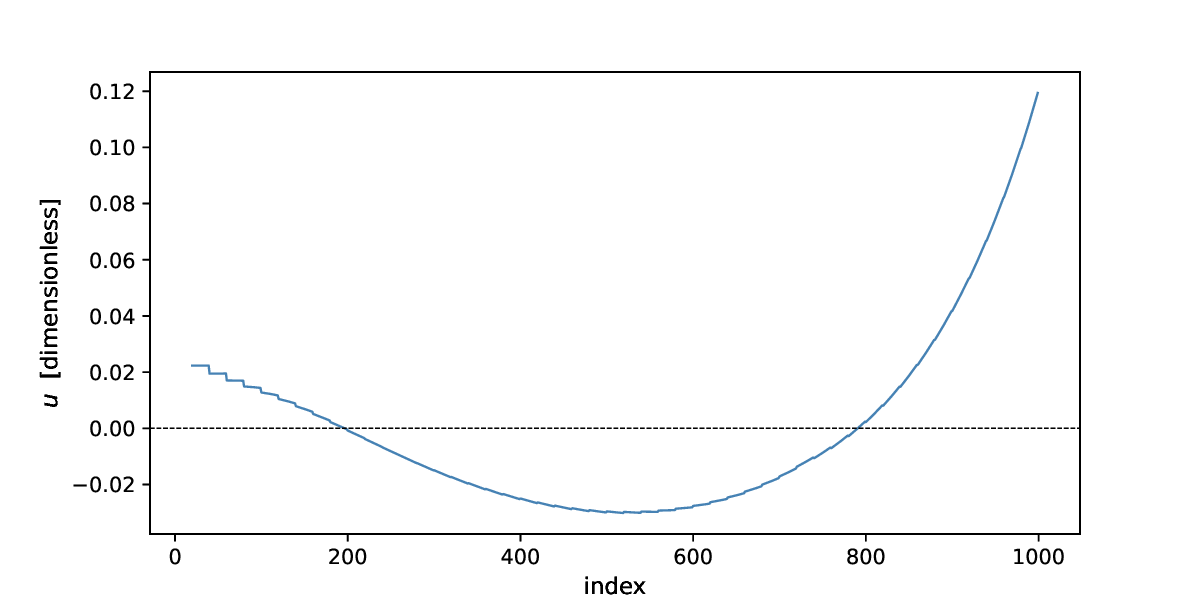}
    \caption{The computed normalized perturbation template $u(t)$. Impulse amplitudes are modulated by the inverse square of the star's radial coordinate $r_1^{-2}$. The template is dimensionless after normalization and x-axis shows the index of the time-step.}
    \label{fig:template}
\end{figure}

Expression (\ref{eq:u}) is the final template, which can be substituted into the formulae above to determine the SNR and thereby detect or rule out the presence of the perturbation in the data. The computed template $u(t)$ is shown in Fig. \ref{fig:template}. 

Finally, in the later section we will compare the delta-function template derived with finite-width pulse model and show that the detection efficiency is stable under reasonable changes in pulse width, cadence, and phase coverage.

\subsubsection{Construction of the SNR statistic}
\noindent

Let the observed series $\rho_{\rm obs}^2(t_k)\equiv\text{SNR}(t_k)$, $k=0,\dots,N-1$, be obtained after applying the optimal filter, with time step $\Delta t$. Here $\text{SNR}(t_k)$  is defined by expression (\ref{eq:SNR}) evaluated at an arbitrary time $t_k$.

We define the integral statistic:
\[
S[\rho]\equiv\int_0^T |\rho^2(t)|\,dt
\approx \sum_{k=0}^{N-1} |\rho^2(t_k)|\,\Delta t .
\]

For the observed line-of-sight velocity series, the statistic of interest is:
\[
S_{\rm obs} = S[\rho^2_{\rm obs}].
\]

We first construct a pure noise model (null hypothesis). To this end, generate $N_{\rm trials}=10^5$ independent noise realizations and apply the same filter to each, obtaining the series $\rho^2_j(t_k)$, $j=1,\dots,N_\text{trials}$.

For each realization we then compute:
\[
S_j = \sum_{k=0}^{N-1} |\rho^2_j(t_k)|\,\Delta t,
\]
thereby defining the empirical noise distribution function:
\[
\widehat F_0(s) = \frac{1}{N_\text{trials}}\sum_{j=1}^{N_{\rm trials}} \mathbf{1}\{S_j \le s\}.
\]

Since the distribution of $S_j$ is non-Gaussian, we adopt a robust estimate based on the median absolute deviation (MAD):
\[
\mathrm{med}_S = \operatorname{median}_{1\le j\le N_\text{trials}}(S_j),
\qquad
\mathrm{MAD}_S = \operatorname{median}_{1\le j\le N_\text{trials}} \bigl| S_j - \mathrm{med}_S \bigr|.
\]

The robust standard deviation estimate is then \cite{Ronchetti}:
\[
\sigma_{\rm robust} = 1.4826 \cdot \mathrm{MAD}_S .
\]

This quantity serves as an analogue of the standard deviation for the noise distribution of the statistic $S$. 

The normalized filter response is defined as
\begin{equation}
\nonumber
Z_{\rm obs}=\frac{S_{\rm obs}}{\sigma_{\rm robust}}.
\end{equation}
Here $\sigma_{\rm robust}$ characterizes the scale of the empirical noise-statistic distribution. Accordingly, $Z_{\rm obs}$ is a measure of the filter response and not a direct Gaussian-equivalent sigma significance. This dimensionless statistic measures how many times the observed integral signal ``energy'' exceeds the characteristic scale of noise fluctuations. It will be used as a relative measure of signal significance for evaluating the efficiency of an observational program.

\subsubsection{Robustness tests and comparison to the finite-width template}
\noindent

Representing the perturbation as a sequence of impulses and building a normalized
template from delta functions is mathematically convenient, but requires some additional tests of the applicability and robustness of this model. Matched filtering can be sensitive to the sharpness of the template; therefore, we compare the detection efficiency with that obtained using a finite-width template. In order to do so, we introduce a step-like impulse template which has the following form:
\begin{equation}
\label{eq:u_fw}
u(t) =\frac{1}{W} \sum_{k=0}^{N_{imp}-1}\frac{\left(a_{k}-\bar{a}\right)}{\tau}\Pi \left(\frac{t-t_{k}}{\tau}\right),
\end{equation}
which contains the same integral perturbation as (\ref{eq:u}) but has a finite width instead of a delta-function peak.

We then choose a Gaia BH-like test system, integrate its motion introducing consecutively a delta-like perturbation and a finite-width one, and calculate the ratio of the peak SNR responses \ref{eq:SNR} in these two cases, $\rho^2_\Pi/\rho^2_\delta$, for a set of different template widths from $0.01T$ to $0.1T$, where $T$ is the orbital period of the perturbing object. 

\begin{figure}
    \centering
    \includegraphics[width=0.9\linewidth]{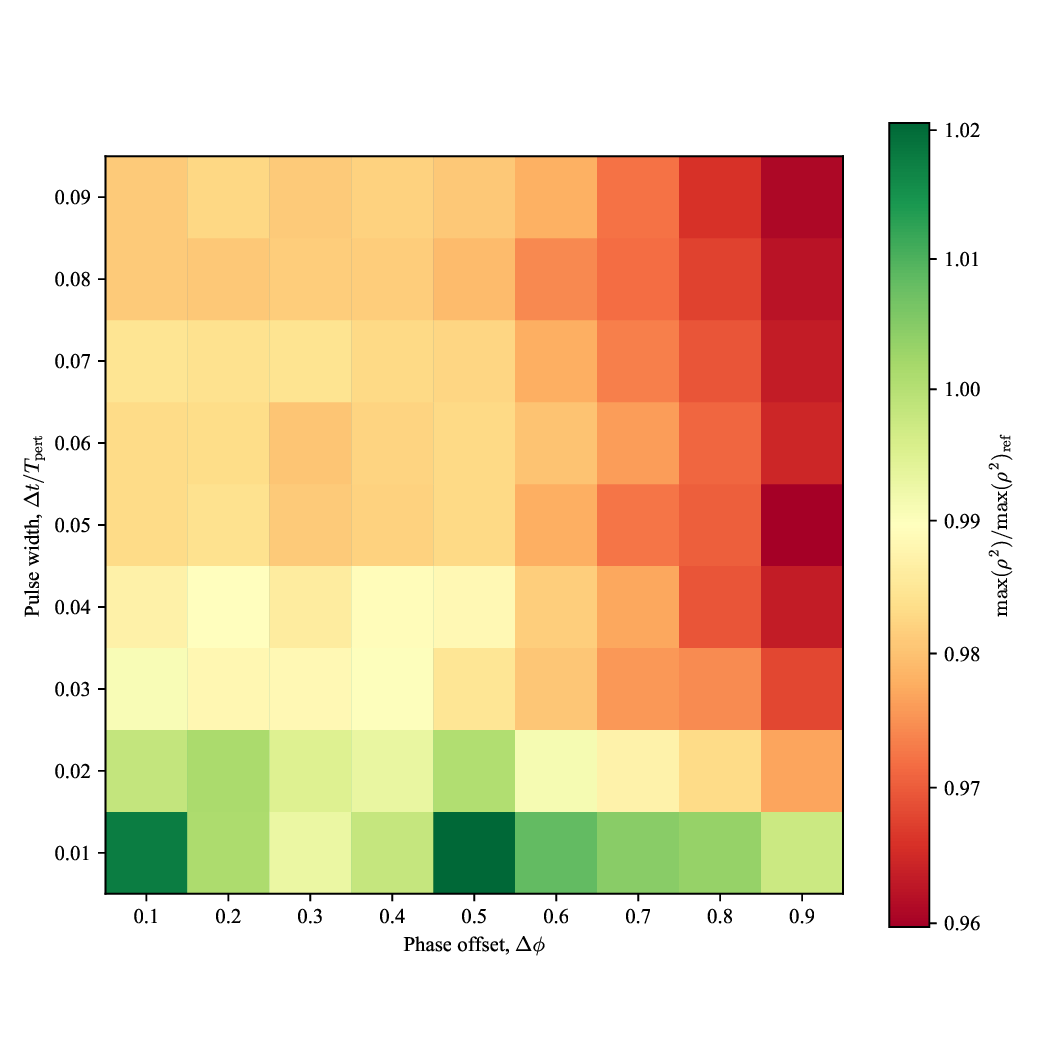}
    \caption{Robustness of the matched-filter response to the pulse width $\Delta t/T_{\rm pert}$ and phase offset $\Delta\phi$ of the perturbation. The color shows the ratio of the maximum response to the reference value obtained for the delta-functional perturbation starting at the pericenter of the test star.}
    \label{fig:robustness}
\end{figure}

Additionally, we study the possible dependence of the response on the orbital phase of the star on side 1 at which the first perturbation is experienced (for simplicity, it was previously considered to be the pericenter with $\phi=0$). For each value of the template, and therefore perturbation, width $\Delta t/T$ and for each value of the phase offset from the pericenter $\Delta \phi$ of the test star on side 1, we integrated its motion and calculated the value of $\frac{\rho^2_\Pi(\Delta t,\phi)}{\rho^2_\delta(0,0)}$. The resulting robustness map is shown in Fig. \ref{fig:robustness}.

One can see that the resulting change in the peak response does not exceed 4\%, with the response at some chosen values of the phase offset and pulse width being even better than for the simplest choice of delta-functional impulses starting from the pericenter of the test star. This allows us to further use the earlier derived simplistic model for the detailed simulations and building an observational program.

\section{Simulation of the perturbing object's effect on the line-of-sight velocity of the observed star}
\noindent

We now apply the framework to the real system Gaia BH1, for which the expected perturbing acceleration was computed previously in \cite{Moiseev 1}, and provide estimates and features of the simulated line-of-sight velocity perturbation profile. Results for synthetic systems are also presented.

\subsection{Radial velocity perturbations simulated for Gaia BH1}
\noindent

For a clearer illustration of the effect, we artificially amplify the perturbing acceleration in the equations of motion by a factor of $10^3$ relative to the maximum estimate for Gaia BH1 (the previously obtained perturbing acceleration is of order $10^{-4}$ cm/s$^2$, \cite{Moiseev 1}). The remaining parameters (mass of the dark body,  mass of the companion star, eccentricity, and period) are taken from the real system: $M=9.62 \pm 0.18 M_{\odot}$, $m=0.93 \pm 0.05 M_{\odot}$, $e=0.451 \pm 0.005$, $P=185.59 \pm 0.05$ days.

\begin{figure}
    \centering
    \includegraphics[width=1\linewidth]{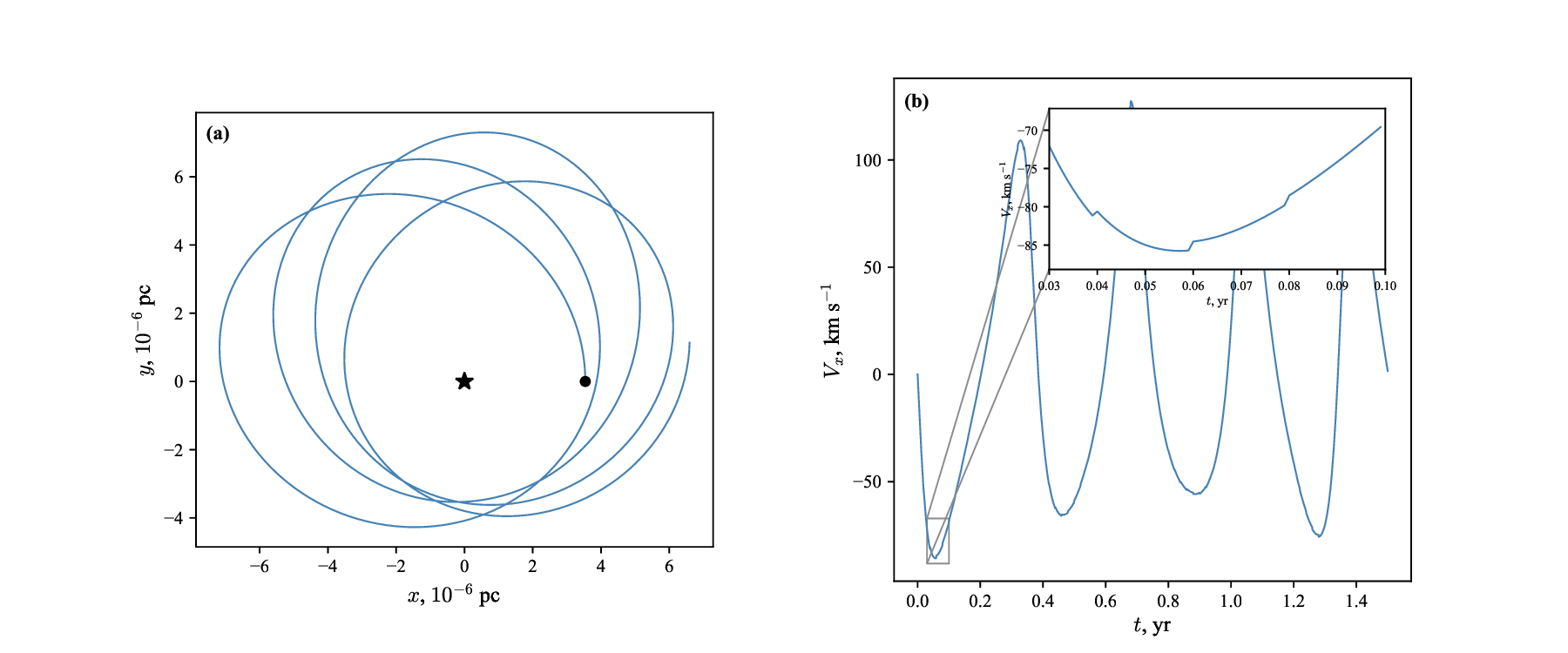}
    \caption{\textit{Left}: integrated trajectory of the star in the orbital plane. \textit{Right}: $x$-component of the stellar velocity. The inset demonstrates how small the sought perturbations are even when artificially amplified by a factor of 1000 in the simulation.}
    \label{fig:exaggerated}
\end{figure}

It is important to note (Fig. \ref{fig:exaggerated}) that even at a significantly amplified scale, the perturbation impulses themselves are invisible against the background of the observed star's motion.

\begin{figure}[h!]
    \centering
    \includegraphics[width=1\linewidth]{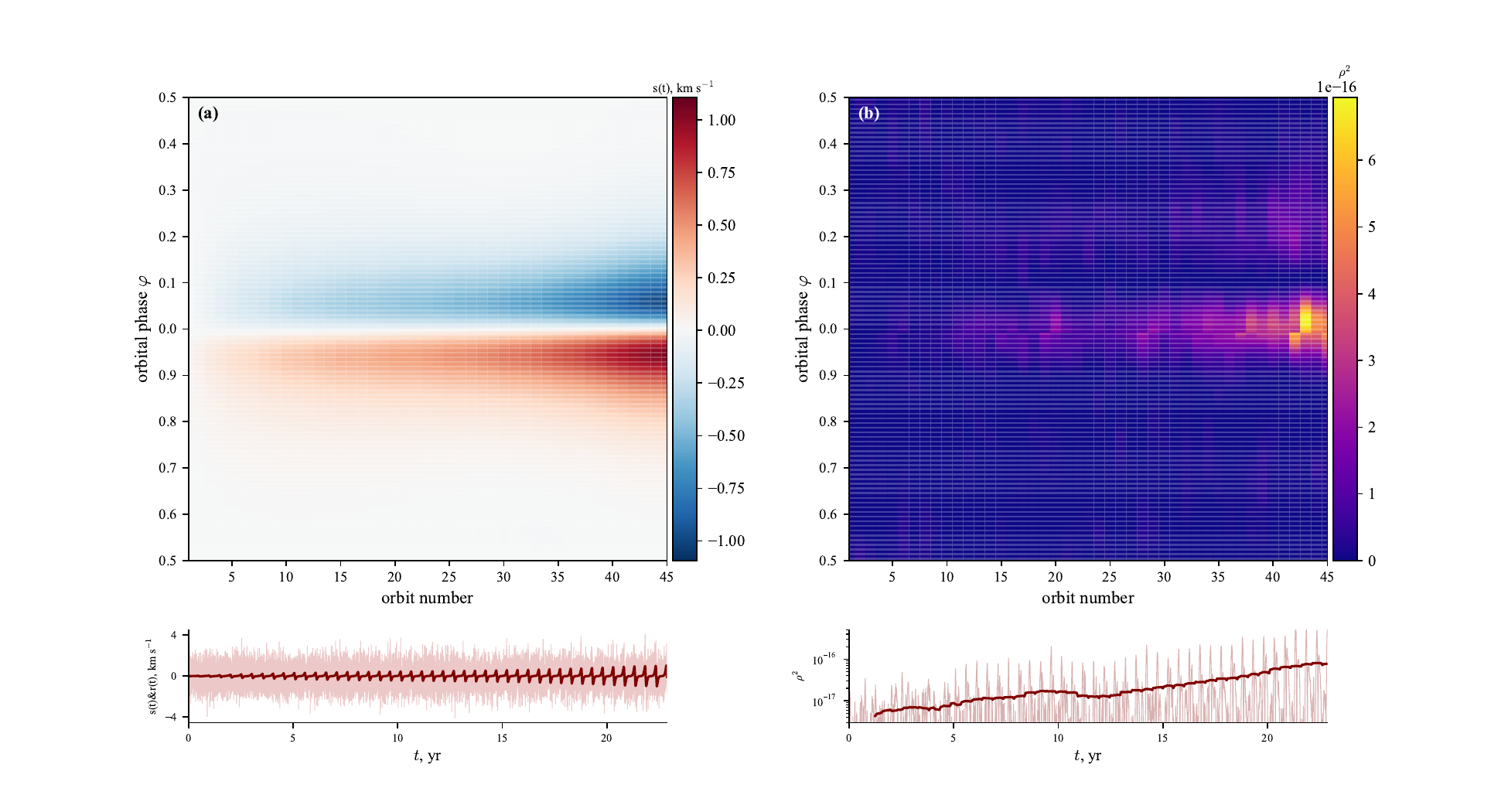}
    \caption{\textit{a}) Top: the ideal signal $s(t)=v^{o}-v_{0}$ (where $v_{0}$ is the Keplerian component and $v^{o}$ is the observed line-of-sight velocity of the star) as a function of orbital phase and orbit number; bottom: $s(t)$ as a function of observation time, with the $\pm 1$ km/s observational noise shown in the background. \textit{b}) Top: the matched filter statistic $\rho^2$ for the noisy observed signal in the phase-domain representation; bottom: in the time-domain representation, with the running mean overlaid. Note that $\rho^2$ is dimensionless; its absolute value depends on the internal normalization of the template and is meaningful only relative to its empirical noise distribution.}
    \label{fig:filtering_res}
\end{figure}

We now adopt the realistic perturbing acceleration of 10$^{-4}$cm/s$^2$ computed previously. Fig. \ref{fig:filtering_res} summarizes the simulation results: the ideal signal and the picture of its growth with time in both the phase and time domains, together with the matched filter statistic $Z_{\text{obs}}$ constructed from the noisy data.

As is evident from Fig. \ref{fig:filtering_res}, the dominant contribution to the perturbation $s(t)$ comes from the shortening of the orbital period rather than from the individual perturbation impulses. The expanding pattern clearly visible in the lower panel is the response to the period decrease, against which the individual impulses are invisible. In the phase-domain representation, the sharp rise and fall of $s(t)$  near pericenter at phase $\phi = 0$ are clearly visible. The matched filter statistic shows a prominent and growing maximum near pericenter, corresponding to the largest perturbation contribution.

In the case of real observations, the data naturally form a discrete and, in general, non-uniform set. We retain the same perturbation amplitude for Gaia BH1 $\sim 10^{-4}$ cm/s$^2$, and consider a hypothetical discrete observational dataset of 100 radial velocity measurements per year.

\begin{figure}[h!]
    \centering
    \includegraphics[width=1\linewidth]{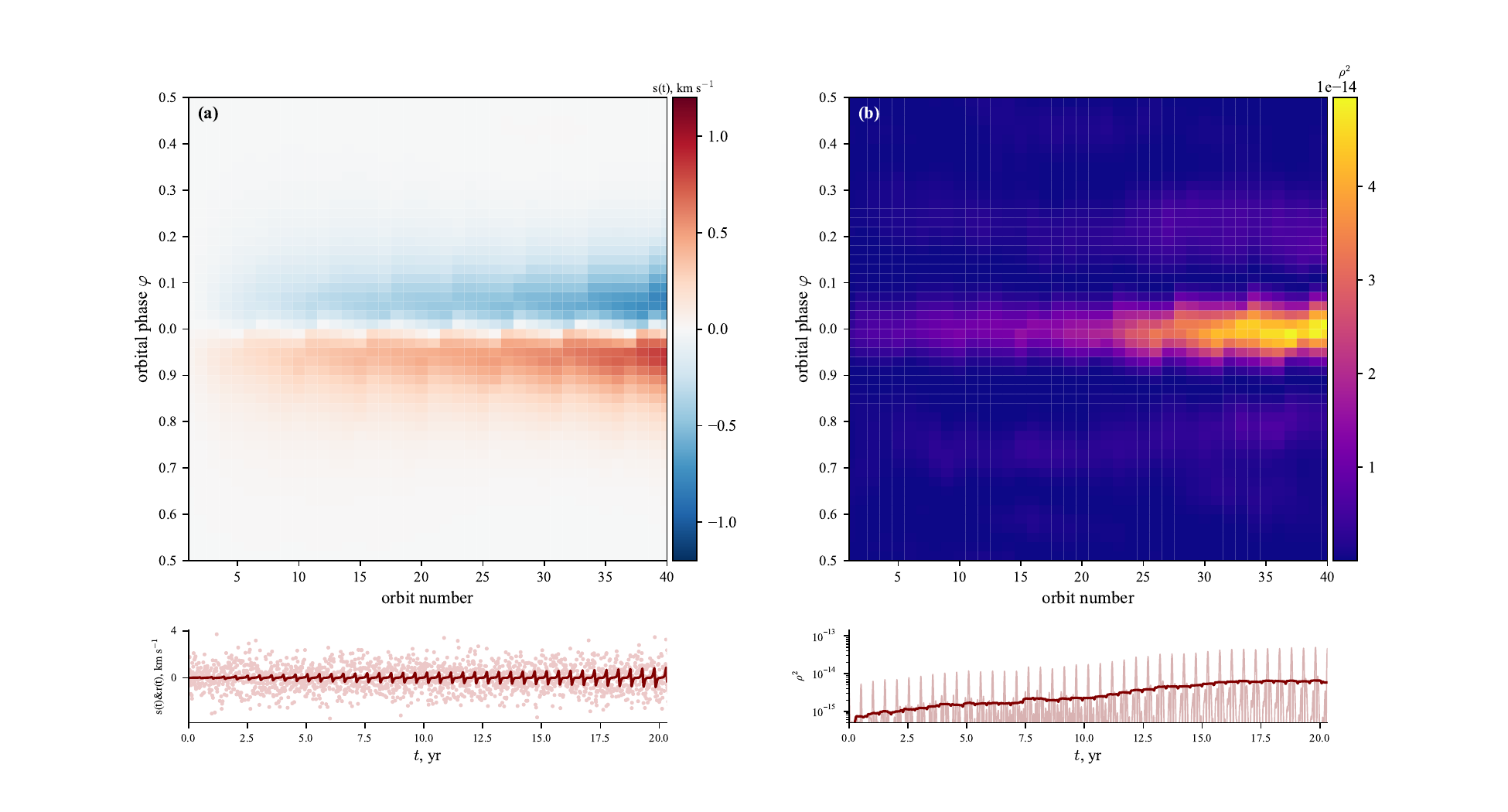}
    \caption{\textit{a}) Top: discrete noiseless observations in the phase-domain representation; bottom: the discrete time series, with the $\pm 0.5$ km/s.  \textit{b}) Top: the matched filter statistic $\rho^2$  for the noisy discrete observed signal in the phase-domain representation; bottom: in the time-domain representation, with the running mean overlaid. As in Fig. \ref{fig:filtering_res} $\rho^2$ is dimensionless and interpretable only in comparison to its empirical noise distribution.}
    \label{fig:discrete_data}
\end{figure}

As seen in Fig. \ref{fig:discrete_data}, even for such a comparatively small number of data points, the matched filter response remains pronounced and shows the perturbation near pericenter.

The following conclusions can be drawn from the results obtained.
\begin{itemize}
\item A long radial velocity time series of the companion star orbiting the WH candidate is required; the necessary observation baseline should be refined through simulations accounting for current and expected observational precisions.
\item Complete observational coverage is not strictly necessary, since the effect manifests itself even in a comparatively sparse discrete dataset of 100 points, provided the observation baseline is sufficiently long. Nevertheless, uniform sampling is preferable, as non-uniform observations degrade the filter response.
\item The dominant contribution to the sought effect comes from the shortening of the orbital period. The jump in the residual $v^{o}-v_{0}$ is not due to the perturbation impulse itself, but to the lag between the predicted and observed velocity maximum.
\end{itemize}

Finally, we discuss the significance of the period change of the companion star. Following \cite{Simonetti}, the formula for the change in the observed star's period per perturbation impulse is:
\[
\delta P\sim6\pi\frac{m}{M}\frac{r_g}{r_p}f^2T,
\]
where $r_g$ is the gravitational radius of the dark object with mass $M$, $f=r_p/r_a$, $m$ is the mass of the perturbing object. 

This leads to a total period change over the observation baseline $\Delta T$
\[
\Delta P\sim\delta P\frac{\Delta T}{P}.
\]

In our simulations this effect is most conveniently determined numerically by tracking the moments of pericenter passage. The shift in the pericenter passage time is found to be at the level of $\sim 10^{-6} - 10^{-5}$ yr, which is one to two orders of magnitude below the current precision of period determination from orbital solutions. Consequently, the period change cannot be estimated with sufficient accuracy from current orbital solutions.

\subsection{Observational strategy and its efficiency}
\noindent

The conclusions above can be confirmed by a numerical analysis of the efficiency of various observational strategies for detecting radial velocity perturbations of the companion star, using a set of synthetic systems.

We assign random parameters over wide ranges (permitting significant radial velocity perturbations). For computational convenience we fix not the mass and orbital parameters of the perturbing object on side 2, but rather the perturbation amplitude $C$ from formula (\ref{eq:deltav}):

\begin{itemize}
    \item $M = M_{\text{WH}}=5-50M_{\odot}$;
    \item $e_1 = 0-0.95$ (eccentricity of the companion star);
    \item $C = 10^{-22}-5\cdot10^{-20}$;
    \item $i = 0^{\circ}-180^{\circ}$
    \item $\Omega = 0^{\circ}-360^{\circ}$
\end{itemize}

\begin{figure}[h!]
    \centering
    \includegraphics[width=0.9\linewidth]{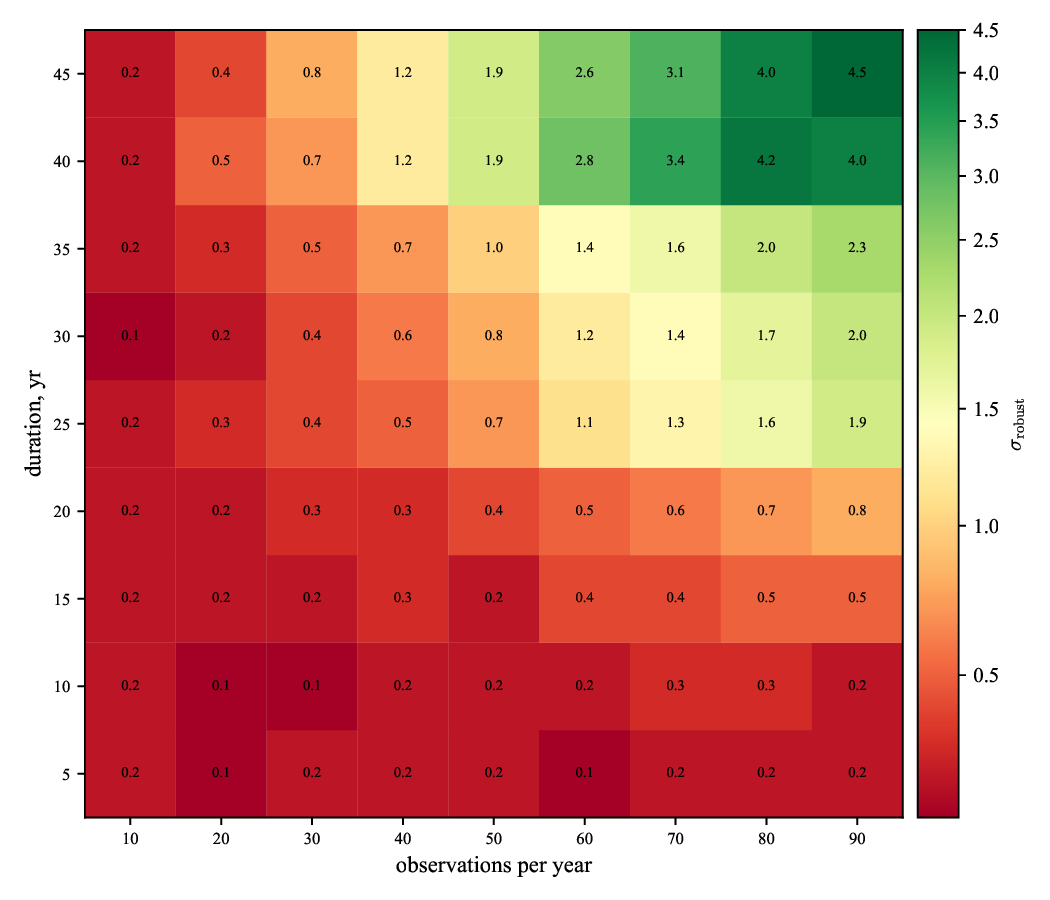}
    \caption{Averaged detection efficiency map based on 500 synthetic systems. The color encodes the mean value of the normalized integral statistic $Z_{\text{obs}}=S_{\text{obs}}/\sigma_{\text{robust}}$  where larger values indicate a stronger normalized filter response. The axes show the observation baseline $T$ and the number of observations per year, illustrating how detection efficiency scales with both the duration and density of the observational program.}
    \label{fig:strategy}
\end{figure}

For each generated system, we integrate the motion over different time baselines and with different numbers of observations per year. For each case we apply matched filtering to the computed radial velocity series and determine $Z_{\text{obs}}=S_{\text{obs}}/\sigma_{\text{robust}}$. The averaged result over 500 randomly built systems is shown in Fig. \ref{fig:strategy}.

The simulation confirms that the longer the observation baseline, the more reliably the sought signal can be identified.

\section{Spectral and non-parametric analysis of the model}
\noindent

The matched-filter amplitude underlying the normalized integral statistic $Z_{\rm obs}$ projects the template onto the observed
signal and weights the result with the template norm:
\[
\rho=\frac{\langle r|u\rangle}{\sqrt{\langle u|u\rangle}}.
\]
meaning that it operates without accounting for the expected periodic nature of the perturbation, suppressing the noise in the frequency domain (via the $1/S_n$ weight in (\ref{eq:SNR})) and marking the segments of the signal that resemble the model perturbation the most.

Additionally, we should study the problem in the spectral formulation, using the LS periodogram as a diagnostic tool for identifying the expected periodicity in the observed time series $r(t)=v^{o}(t)-v_{0}(t)$, where $v_{0}(t)$ -- the Keplerian component, $v^{o}(t)$ -- the observed line-of-sight velocity of the star as functions of time.

\subsection{LS periodogram}
\noindent

The problem of a periodic structure search in the noise-dominated and potentially non-uniform observational time series $r(t)=v^{o}(t)-v_{0}(t)$ can be solved with the LS periodogram \cite{T}, which has the following form:
\[
P_N(\nu)
=
\frac{
(2 - C_N) U_N^2
+
C_N V_N^2
-
2 S_N U_N V_N
}{
C_N (2 - C_N) - S_N^2
},
\]
where
\[
C_N(\nu) = \frac{2}{N} \sum_{k=0}^{N-1}
\cos^2 \left( 2\pi \nu t_k \right),
\]
\[
S_N(\nu) = \frac{1}{N} \sum_{k=0}^{N-1}
\sin \left( 4\pi \nu t_k \right),
\]
\[
U_N(\nu) = \frac{1}{N} \sum_{k=0}^{N-1}
r_k \cos \left( 2\pi \nu t_k \right),
\]
\[
V_N(\nu) = \frac{1}{N} \sum_{k=0}^{N-1}
r_k \sin \left( 2\pi \nu t_k \right).
\]

The numerical evaluation of these formulae can be done over any frequency grid in a potentially relevant range $\nu=[\nu_{\text{min}}\,;\nu_{\text{max}}]$.

\begin{figure}
    \centering
    \includegraphics[width=0.7\linewidth]{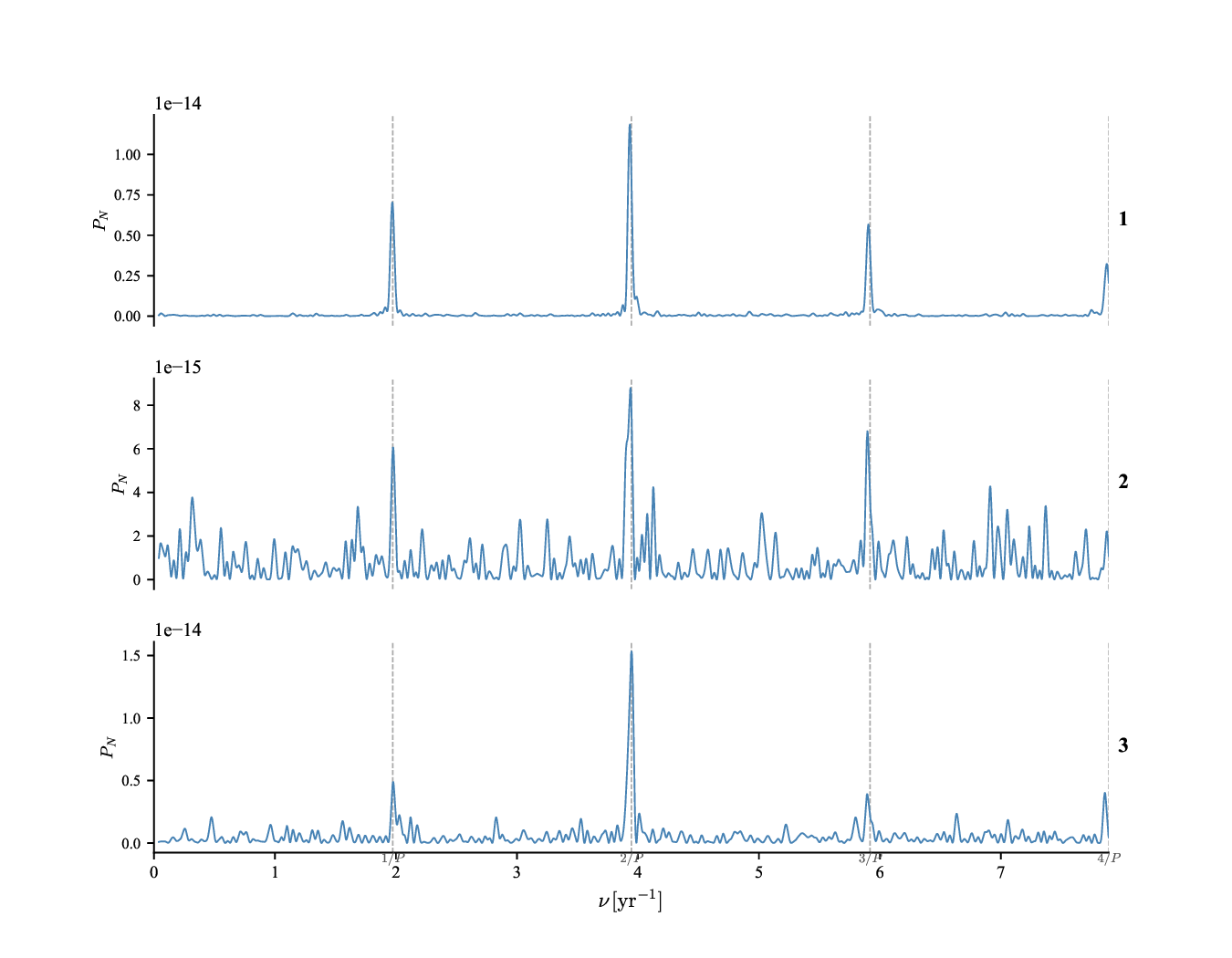}
    \caption{LS periodograms of $r(t)$. Panel 1: densely sampled uniform series, noise 1 km/s; panel 2: densely sampled uniform series, noise 4 km/s; panel 3: sparsely and non-uniformly sampled series, 100 observations per year, noise 1 km/s. Dotted vertical lines mark the expected $k/P$ harmonics.}
    \label{fig:all_LS}
\end{figure}

In Fig. \ref{fig:all_LS} (panel 1) we present the LS periodogram computed for a densely sampled 15-year time series $r(t)$  with the noise level of $1$ km/s. One can clearly see the $k/P$ harmonics, which correspond to the expected perturbations of the difference between the observed and predicted Keplerian radial velocities near pericenter passage due to the shortening of the system's orbital period.

For a higher noise level of 4 km/s, the result for the densely sampled time series is shown in panel 2 of Fig. \ref{fig:all_LS}. The main harmonics still dominate, although the relative noise level has increased.

The periodograms above were computed for a densely sampled uniform time series with $10^3$ observations per year (15000 points in total). Let us decrease the number of observations. In panel 3 of Fig. \ref{fig:all_LS} there is the result for 100 observations per year with the noise level of 1 km/s with observations distributed non-uniformly across the time series. As one can see in the figure, even in such a case the structure of the peaks does not fully wash out. Therefore, the presence of a harmonic response can serve as an independent indicator of the sought perturbation in the signal.

\begin{figure}
    \centering
    \includegraphics[width=0.7\linewidth]{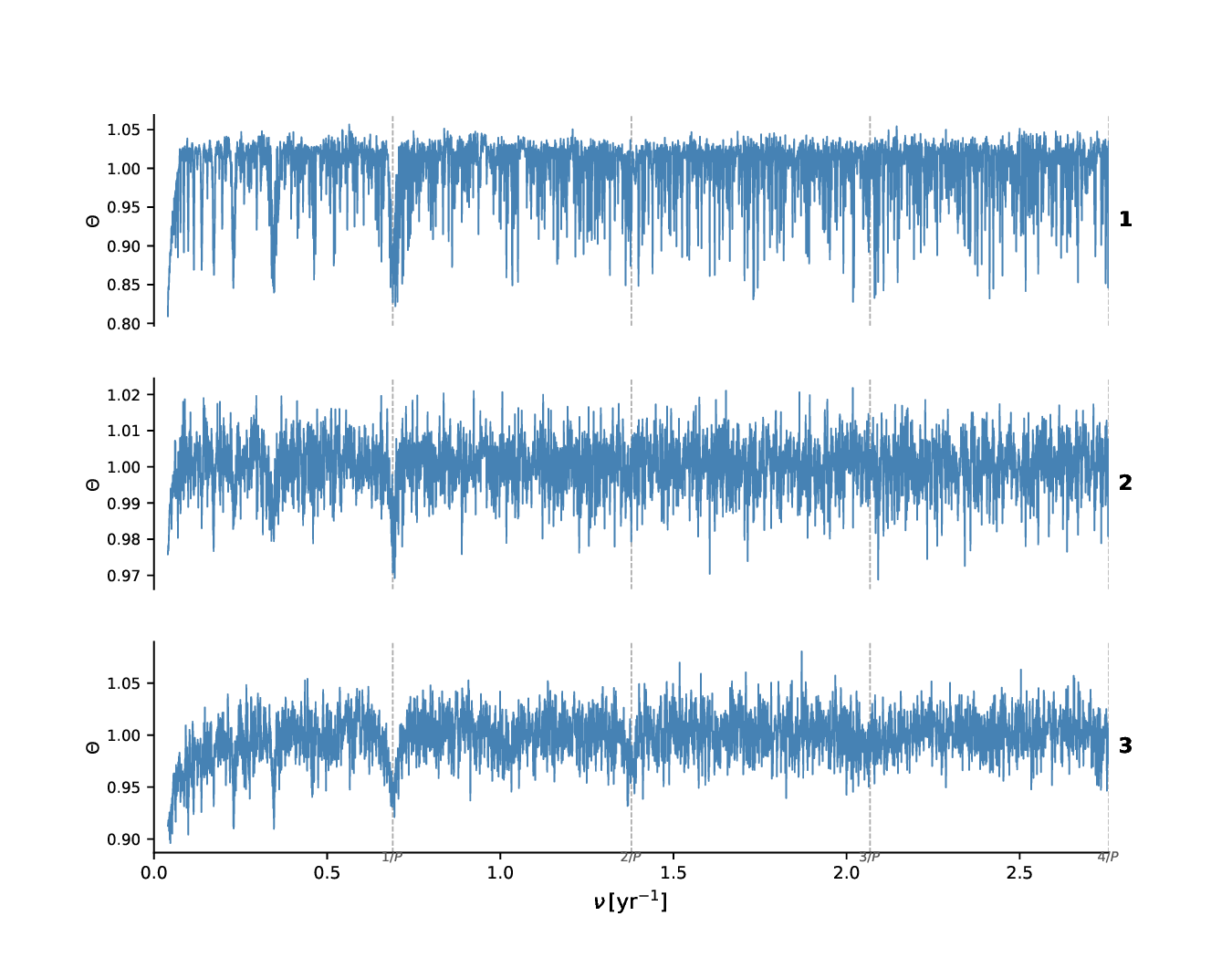}
    \caption{ALK statistic computed for $r(t)$. Panel 1: densely sampled uniform series, noise 300 m/s; panel 2: densely sampled uniform series, noise 1 km/s; panel 3: sparsely and non-uniformly sampled series, 100 observations per year, noise 0.5 km/s. Dotted vertical lines mark the expected $k/P$ harmonics.}
    \label{fig:ALK}
\end{figure}

\subsection{The Abbe--Lafler--Kinman statistic}
\noindent

In addition to the spectral analysis, we employ a non-parametric method based on the Abbe–Lafler–Kinman (ALK) statistic.

For a given time series $y_k = r(t_k)$ and a trial frequency $\nu$, the values are sorted by phase:
\[
\phi_k = \{\nu t_k\}, \quad \text{where $\{\cdot\}$ denotes the fractional part},
\]
\[
(\phi_k, y_k) \to (\phi_{(i)}, y_{(i)}),
\]
and the following statistic is computed:
\[
\Theta(\nu) = \frac{\sum_{i=0}^{N-2} (y_{(i+1)} - y_{(i)})^2}{2 \sum_{i=0}^{N-1} (y_{(i)} - \bar y)^2}.
\]

Minima of $\Theta(\nu)$ correspond to candidate periodicities. For a random time series, $\Theta(\nu)$ is expected to fluctuate around unity \cite{T}.

The results of the ALK statistic for time series with different noise levels, ranging from 300 m/s to 1 km/s, are shown in Fig. \ref{fig:ALK}.

As shown in Fig. \ref{fig:ALK}, for our model of an impulsive perturbation in the presence of background noise, this statistic is less efficient than the LS periodogram, likely due to local fluctuations that can exceed the signal at the $k/P$ harmonics. However, the first peak, corresponding to the main harmonic, remains statistically significant against the noise background, as follows from the efficiency estimates of both methods presented below.

\begin{figure}
    \centering
    \includegraphics[width=1\linewidth]{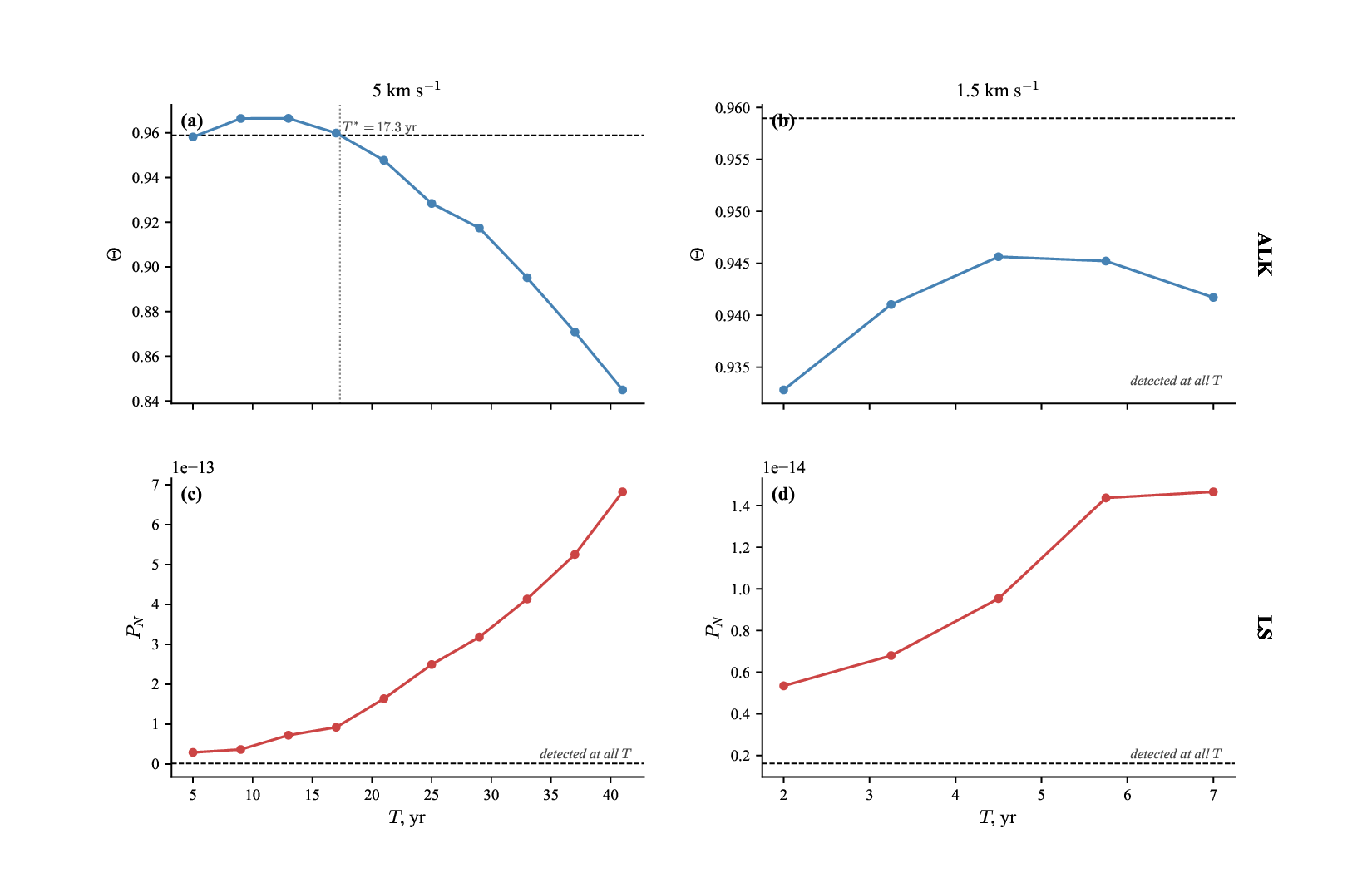}
    \caption{Detection efficiency of the ALK (top row) and LS (bottom row) statistics as functions of observation baseline $T$ for two noise levels. Panel (a): ALK statistic at 5 km/s noise; the vertical dotted line marks $T^*=17.3$ yr, the minimum observation baseline required for the statistic to reach significance. Panel (b): ALK statistic at 1.5 km/s noise; the statistic exceeds the detection threshold at all considered values of $T$. Panels (c) and (d): LS periodogram peak power at 5 km/s and 1.5 km/s noise, respectively; the signal likewise exceeds the threshold at all considered values of $T$ in both cases. Horizontal dashed lines mark the critical values derived from the empirical distributions of each statistic.}
    \label{fig:placeholder}
\end{figure}

Overall, spectral and non-parametric analyses prove to be powerful tools for confirming the presence of the sought perturbations in radial velocity time series. Provided that the time series is sufficiently long and well sampled, the anomalous acceleration effect leaves an imprint in the difference $r(t) = v^{o}(t) - v_{0}(t)$. This imprint can be traced both using matched filtering techniques and through spectral and non-parametric analyses in the frequency domain, showing the harmonic structure induced by the shift of the pericenter passage.

The LS and ALK thresholds were estimated from $10^5$ noise realizations using a nominal false-alarm probability of $q=10^{-5}$. The same frequency search within $\nu_{\rm test}\pm1/T$ was applied to the signal and noise data, selecting the maximum LS power and minimum ALK statistic. This probability is at the Monte Carlo resolution limit and should be regarded as approximate.

In Fig.~\ref{fig:placeholder} we present the resulting efficiency of the search for the sought radial velocity perturbation using the LS periodogram and the ALK statistic. For the observational accuracy in radial velocity expected in the near future, at the level of 1.5 km/s the perturbations are detectable for all considered observation time spans. For a more realistic accuracy of 5 km/s, the ALK method becomes statistically significant after $\approx17$ years of data accumulation. The application of the methods considered in this chapter significantly decreases the required accumulation time.

\section{The effect of uncertainties in the orbital parameters}

As shown in previous sections, assuming that the ordinary Keplerian orbit can be subtracted, the remaining WH signal can then be searched for. However, since the main manifestation of the perturbation is the gradual change in the orbital period and the shift in the time of the velocity maximum, it becomes important to check if the uncertainties in the Keplerian parameters can mask the presence of the signal. 

In order to make a basic estimate of the effect of uncertainties in the Keplerian parameters, a simple numerical test was conducted. We generated simulated data containing the proposed WH signal and then fitted the data in two ways: using a normal Keplerian orbit only, and using a Keplerian orbit together with the WH signal. The orbital parameters and the systemic radial velocity were allowed to change during the fit. The parameters of the test are listed in Table~\ref{tab:kepler_fit_setup}. This test shows whether the WH signal can be separated from orbital-parameter uncertainties under the adopted observing conditions.

\begin{table}[h!]
\centering
\caption{
Parameters of the numerical experiment. The same observation times (5000 points spread uniformly over a 25-year timespan) and Gaussian noise realization (with $\sigma_v=0.1\,\mathrm{km\,s^{-1}}$) are used for all the perturbation amplitudes considered.
}
\label{tab:kepler_fit_setup}
\begin{tabular}{lll}
\toprule
Parameter &  & Value \\
\midrule
Central mass
    & $M$ & $9.62\,M_\odot$ \\
Initial orbital period
    & $P_0$ & $185.59\,\mathrm{d}$ \\
Initial eccentricity
    & $e_0$ & $0.451$ \\
Initial RV semi-amplitude
    & $K_0$ & $86.53462\,\mathrm{km\,s^{-1}}$ \\
Initial radial-velocity offset
    & $\gamma_0$ & $0\,\mathrm{km\,s^{-1}}$ \\
Argument of pericenter
    & $\omega_0$ & $126.6^\circ$ \\
Initial mean anomaly
    & $\mathcal{M}_0$ & $0$ \\
Observation window
    & $T_{\rm obs}$ & $25\,\mathrm{yr}$ \\
Number of measurements
    & $N$ & $5000$ \\
RV uncertainty
    & $\sigma_v$ & $0.1\,\mathrm{km\,s^{-1}}$ \\
Impulse period
    & $T_{\rm imp}$ & $0.02\,\mathrm{yr}$ \\
Perturbation amplitude
    & $C_{\rm true}$
    & $\{10^{-21},10^{-19},10^{-17}\}\,\mathrm{pc^3\,yr^{-1}}$ \\
\bottomrule
\end{tabular}
\end{table}

The Keplerian radial-velocity model is
\begin{equation}
\nonumber
v_{\rm K}(t)=\gamma+K\left[\cos\bigl(\nu(t)+\omega\bigr)
+e\cos\omega\right],
\end{equation}
where $\nu(t)$ is the true anomaly, determined by $P$, $e$,
and the initial mean anomaly $\mathcal{M}_0$; $K$ is the
radial-velocity semi-amplitude, $\omega$ is the argument of
pericenter, and $\gamma$ is the constant radial-velocity offset.
The WH model is obtained by integrating the perturbed equations
of motion and reduces to the Keplerian model at $C=0$.

The results are shown in table \ref{tab:kepler_fit_results}. For the two smaller amplitudes, neither K+WH optimization run improves upon the nested $C=0$ solution, which is therefore retained.
The difference between the corresponding reduced statistics arises solely from the different numbers of fitted parameters.

\begin{table}[htbp]
\centering
\caption{
Comparison of the Keplerian-only (K) and perturbed (K+WH) fits. The parameters $P$, $e$, $K$, $\omega$, $\mathcal{M}_0$, and the systemic radial velocity $\gamma$ are free in both models; the K+WH model additionally allows $C\geq0$. The central mass, impulse period, and impulse phase are fixed. The reduced statistics are $\chi^2_{\nu,{\rm K}}=\chi^2_{\rm K}/(N-6)$ and $\chi^2_{\nu,{\rm K+WH}}=\chi^2_{\rm K+WH}/(N-7)$, with $\Delta\chi^2=\chi^2_{\rm K}-\chi^2_{\rm K+WH}$.}
\label{tab:kepler_fit_results}
\begin{tabular}{cccccc}
\toprule
$C_{\rm true}$
& $\chi^2_{\nu,{\rm K}}$
& $\chi^2_{\nu,{\rm K+WH}}$
& $\Delta\chi^2$
& ${\rm RMS}_{\rm K}$
& ${\rm RMS}_{\rm K+WH}$ \\
$[\mathrm{pc^3\,yr^{-1}}]$
& & &
& $[\mathrm{km\,s^{-1}}]$
& $[\mathrm{km\,s^{-1}}]$ \\
\midrule
$10^{-21}$ & 0.9837 & 0.9839 & 0.00
    & 0.09912 & 0.09912 \\
$10^{-19}$ & 0.9835 & 0.9837 & 0.00
    & 0.09911 & 0.09911 \\
$10^{-17}$ & 1.5091 & 0.9838 & 2624.35
    & 0.12277 & 0.09912 \\
\bottomrule
\end{tabular}
\end{table}

One can see that, among the cases tested, only for the extreme perturbation amplitude $C=10^{-17}\,\mathrm{pc^3\,yr^{-1}}$,
which can be realized for a large throat
($\gtrsim10^2$ gravitational radii) and a large perturbing acceleration ($\gtrsim10^{-2}\,\mathrm{cm\,s^{-2}}$), does the Keplerian model fail to produce a sufficiently accurate fit, whereas for smaller perturbations it can absorb the effect of the WH signal. Thus, although the sought effect is identifiable under reasonable assumptions when the underlying Keplerian orbit is known exactly, a free fit of orbital parameters represents a significant limitation for the isolation of the desired signal and must be taken into account when moving from idealized detectability estimates to the analysis of real observations.

\section{Discussions and conclusions}
\noindent

Wide binary systems with a luminous companion star and a compact object with negligible or absent accretion belong to the class of so-called dormant systems: dBH-LC systems, \cite{Liu}--\cite{Wang}, which have been actively studied recently.

In particular, a number of papers discuss the Gaia BH1 pair, whose compact object origin, if it is a stellar-mass BH, encounters a number of difficulties. The problem is that it is impossible to explain the existence of a supergiant that collapsed into a BH, since its estimated radius ($R\sim 10^3 R_{\odot}$) does not correspond to the current orbital parameters of this system, \cite{Hurley2000}. In the case of the compact component, which we assume to be WH, there is no need to consider it as an evolved stellar remnant.

Further, the assumption of the triple nature of this system involves two approaches. So, if the third component is located at a sufficient distance from the central compact object, then the problem can be reduced to the dynamics of a triple system, the third component of which is a massive ``planet'' in the orbit of a companion star \cite{Toshinori}. In this case, the perturbation induced by the third body in the radial velocity of the companion may differ in its time dependence from the cumulative shift predicted by our WH model. In the second case, the third companion is close to the compact object, although no evidence for such an inner binary was found in Gaia BH1 \cite{Nagarajan}. If the luminous companion star orbits an inner binary BH (a compact BH+BH system), its radial velocity may contain components at combinations of the orbital frequencies. The two scenarios may therefore have different radial velocity signatures, but a quantitative assessment of their distinguishability is left for future work.

We have shown that the perturbing effect of the radial velocity of a companion star in the case of a traversable WH can be detected in observations of binary wide systems assuming that the underlying Keplerian orbit is known. The WH structure in our model is described by stitching two Schwarzschild metrics, which is a fairly accurate approximation for the wide systems under consideration. The traversability of WH can be provided, for example, by phantom energy. The DESI results provide interesting cosmological hints towards evolving or phantom-like dark energy. Although they do not by themselves confirm the existence of the exotic matter distribution required to stabilize an astrophysical traversable WH, the use of phantom dark energy as a source of the exotic stress-energy required for traversability remains a theoretically motivated possibility, and can be relevant for the considered WH model. We also assume the natural presence of a massive object on side 2, which, being in an elliptical orbit near the WH, produces the sought anomalous acceleration of the companion star on side 1, manifesting as a perturbation in its radial velocity.

The present analysis revealed an important limitation that concerns the accuracy with which the Keplerian component of the motion can be subtracted, i.e., how well the expected radial velocity $v_{0}(t)$ can be predicted. On the one hand, orbital modeling provides us a residual at the level of the instrumental uncertainty. On the other hand, uncertainties in the orbital parameters can degrade our detectability estimates. As shown by the simple numerical test in Section 5, adjustments of the Keplerian parameters can absorb a significant portion of the perturbation. In a realistic observational analysis, such uncertainties and systematics should therefore be considered as an additional contribution to the noise and systematic residuals, rather than assuming that the subtraction of the Keplerian component
produces residuals completely free of orbital-model uncertainties. 

Matched filtering may help distinguish the remaining WH signal through its short-scale structure. For a present signal we expect
\[
S_{\text{obs}}\sim\langle r|u\rangle\sim\langle u|u\rangle,
\]
whereas errors in the orbital parameters yield
\[
S_{\text{obs}}\sim \langle \frac{\partial v}{\partial\theta _i}|u\rangle.
\]

Thus the key to separating the sought effect from parameter errors might be in the use of a local, accurate template, which in our case has strong discrete steps on small scales. Such a template, being sensitive to local jumps in the radial velocity, may not be fully reproduced by the apriori smooth and non-local parameter error. The projection of the template onto the variation $\frac{\partial v}{\partial\theta_i}$ may therefore be small, since the local contributions of the short-scale structure of the template are averaged against the smooth variation of the orbital parameters, while the signal $\langle u|u\rangle$ adds coherently. 

As a simplified order-of-magnitude illustration, one can consider a representative orbital-period uncertainty of $\delta P/P=10^{-5}\div10^{-4}$ and calculate $\Delta v\approx v(t,P+\delta P)-v(t,P)\approx\delta P\cdot\frac{\partial v}{\partial P}$ by integrating the motion of the test system twice and then numerically evaluating the ratio
\[
\frac{\langle u|u\rangle}
{\left|\left\langle\frac{\partial v}{\partial P}\middle|u\right\rangle\right|}\approx\frac{\langle u|u\rangle}{\left|\left\langle\frac{\Delta v}{\delta P}\middle|u\right\rangle\right|}.
\]

For a Gaia BH1-like system, this oversimplified single-configuration estimate gives a value of approximately $1.8\cdot10^{2}\div1.7\cdot10^1$. This result should not be interpreted as a quantitative assessment of the covariance between orbital-parameter uncertainties and the WH-induced perturbation, but rather as an illustration that the short-scale structure of the WH template can substantially reduce its overlap with smooth orbital-model variations. Although the test presented in Section 5 demonstrates the significance of this problem, a more comprehensive assessment of the joint fit, introducing instrument-specific uncertainties,
realistic temporal sampling, and independent constraints on orbital parameters, is needed to establish whether the remaining signal can be separated in practice.

We also took into account several more subtle competing effects, including (1) the influence of the dark matter halo, (2) the mean influence of the background of surrounding stars, and (3) the stochastic gravitational influence of surrounding stars. Among these effects, the stochastic influence of surrounding stars gives the largest estimated contribution and, for S2, can be comparable to or larger than the estimated WH-induced acceleration.

The results obtained in the present work should therefore be understood as model-based estimates of the expected signal and its detectability under the considered assumptions. According to the estimates made, for the accuracy of radial velocity observations of the order of 1.5 km/s the desired effect is detected using spectral (LS periodogram) and nonparametric analysis (ALK method) for all considered observation baselines (based on simulations assuming a minimum of 100 observations per year). At the more conservative accuracy of 5 km/s, representative of current observational limits, the ALK method becomes statistically significant after approximately 17 years of data accumulation. These results show the importance of the considered effect
for a possible WH search in real observational data, with Keplerian parameter uncertainties representing a major obstacle and possible ways to overcome it requiring further detailed study.

\vskip5mm
The research was supported by the Russian Science Foundation, project No. 25-22-20026.

\end{document}